\documentclass{article}
\usepackage{ijcai21}

\usepackage{times}
\usepackage{soul}
\usepackage{url}
\usepackage[hidelinks]{hyperref}
\usepackage[utf8]{inputenc}
\usepackage[small]{caption}
\usepackage{graphicx}
\usepackage{amsmath}
\usepackage{amsfonts,amssymb}
\usepackage{amsthm}
\usepackage{booktabs}
\usepackage{algorithm}
\usepackage{algorithmic}
\title{Dance Generation with Style Embedding: Learning and Transferring Latent Representations of Dance Styles}

\author{
Xinjian Zhang$^1$
\and
Yi Xu$^1$\and
Su Yang$^1$\and
Longwen Gao$^2$ \and
Huyang Sun$^2$
\affiliations
$^1$School of Computer Science, Fudan University, China\\
$^2$Bilibili, China
\emails
\{zhangxj17, yxu17\}@fudan.edu.cn,
\{gaolongwen, sunhuyang\}@bilibili.com
}

\begin{document}

\maketitle

\begin{abstract}
Choreography refers to creation of dance steps and motions for dances according to the latent knowledge in human mind, where the created dance motions are in general style-specific and consistent. So far, such latent style-specific knowledge about dance styles cannot be represented explicitly in human language and has not yet been learned in previous works on music-to-dance generation tasks. In this paper, we propose a novel music-to-dance synthesis framework with controllable style embeddings. These embeddings are learned representations of style-consistent kinematic abstraction of reference dance clips, which act as controllable factors to impose style constraints on dance generation in a latent manner. Thus, the dance styles can be transferred to dance motions by merely modifying the style embeddings. To support this study, we build a large music-to-dance dataset. The qualitative and quantitative evaluations demonstrate the advantage of our proposed framework, as well as the ability of synthesizing diverse styles of dances from identical music via style embeddings.

\end{abstract}

\section{Introduction}

Dancers move elegantly following the rhythm of music. Behind dancing is the hard labor in terms of choreography. Consequently, automatic music-to-dance choreography has emerged as a new topic in multimedia~\cite{lee2019dancing,ye2020choreonet,ren2020self}. Besides, music-to-dance synthesis investigations are vital to machine perception of the latent knowledge regarding dance moves as well as rhythm~(beat) and emotion~(intensity) matching between music and dance, namely, styles. The generated dance motions can be transferred to realistic-like videos by video generation models~\cite{chan2019everybody,ren2020deep} to enable real applications like virtual character generation, games creation and teaching assistant.

\begin{figure}[t]
\begin{center}
	\includegraphics[width=\linewidth]{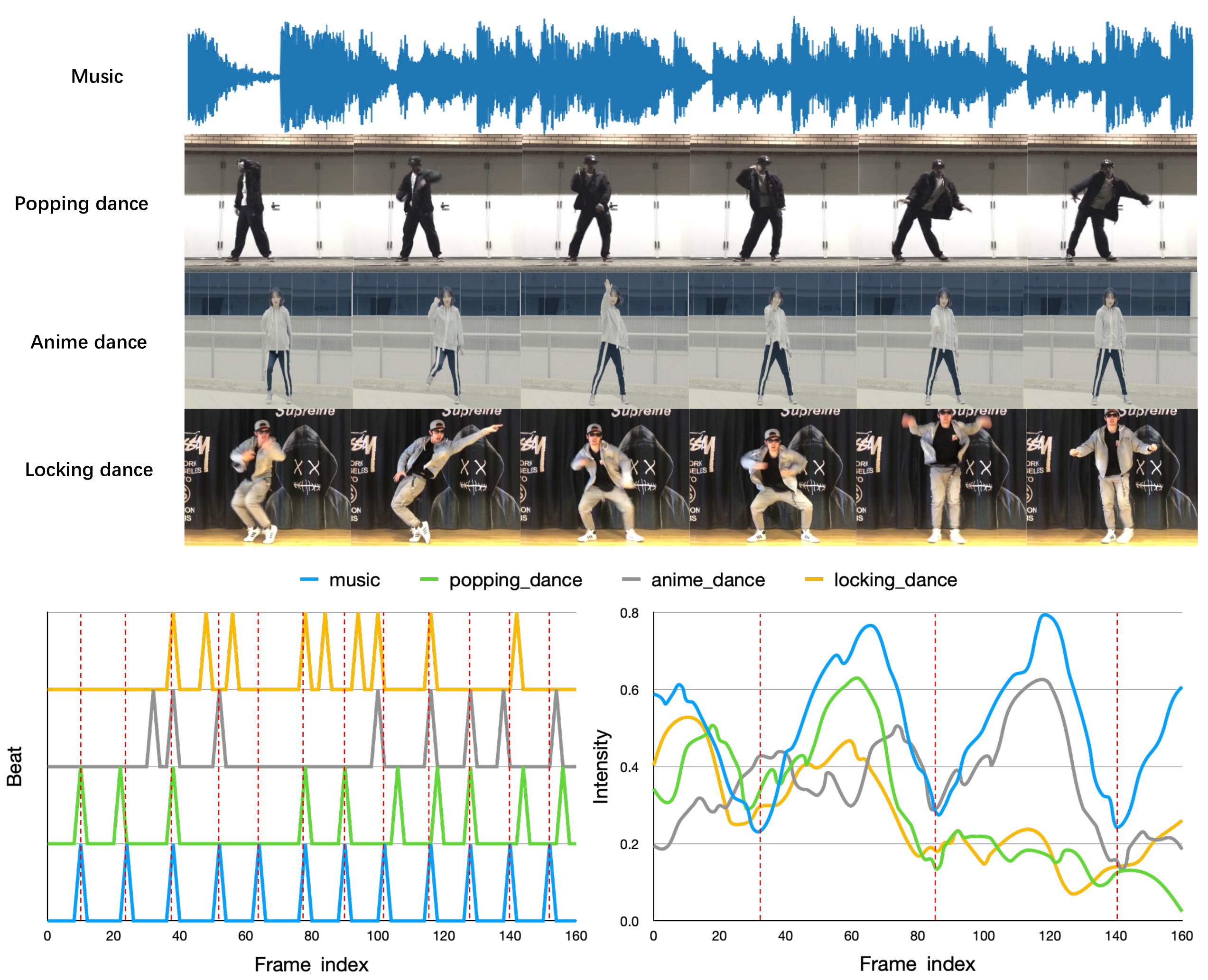}
\end{center}
   \caption{Beat and intensity of popping, anime, and locking dances corresponding to the same music. All dancers' moving rhythm and amplitude match the music well with remarkably different motion patterns in terms of the kinematic representations.}
\label{fig:1}
\end{figure}

Historically, most dance generation methods pay efforts to model the relationship between music and dance in feature space or latent space. Early studies~\cite{shiratori2006dancing,ofli2008audio,fan2011example,asahina2016automatic} are mainly retrieval based, which are focused on selecting the best matching pair from database based on music and dance features. Due to the retrieval nature, these methods cannot generate new dance sequences. In recent years, deep generative models~\cite{alemi2017groovenet,tang2018dance,lee2019dancing,ye2020choreonet,ren2020self} are introduced to alleviate the problem.~\cite{tang2018dance,ren2020self} utilize the temporal indexes and contrastive cost function to promote the model's ability of generation.~\cite{lee2019dancing,ye2020choreonet} attempt to decompose dances into a series of basic dance units and learn to compose a dance by organizing multiple basic dancing units. Yet, machine-generated choreography is far behind human works in terms of diversity because the professional knowledge behind choreography, such as dance styles, is not exploited in previous works.

Human choreographers with different dance experience can compose various styles of dances. As shown in Figure~\ref{fig:1}, the dance moves accompanying the same music are composed of sharp tuning with diverse-style variations when matched to the music in terms of both beat and intensity. Moreover, the advanced choreographer can also learn a particular dance style from a short dance clip and apply it to future dance creation. So far, the mechanism of human perception is not explicit. In this context, style control and transfer in dance generation has been remaining an open problem yet.

To generate dances with controllable styles, we introduce the controllable style embeddings into our dance generation framework. The whole dance generation framework contains two modules: A style embedding producer and a music-to-dance generator. The style embedding producer learns the style information  in a latent manner from motion semantics of reference dance clips so as to output style embedding. The music-to-dance generator contains a music encoder and a dance generator. The former is formulated as a set of transformer encoders~\cite{vaswani2017attention} to obtain semantic and temporal representation from the input music. The latter incorporates both the music representation generated by the music encoder and the style embedding as input to produce outputs the generated dance. Considering that the relationship between music and dance is frame-aligned, the framework generates dance motions in a frame-by-frame way. Benefitting from style embeddings, our framework is capable of creating different styles of dances accompanying the same music by modifying style embedding. 

To support this study, we construct a large music-to-dance dataset with style annotation for training and evaluation. We select three modern dancing categories: Anime dance, popping dance, and locking dance to build the dataset, which have a higher degree of freedom in terms of choreography. We compare our framework with several baselines from the following perspectives: Beat matching, emotion matching, style consistency, and diversity. Besides, we visualize the choreography by using a realistic video generation model~\cite{wang2018pix2pixHD,chan2019everybody} for a better qualititative evaluation.

Our contributions are summarized as follows: (1)~We first propose a novel frame-by-frame dance generation framework with controllable style embedding to guarantee style consistence and allow flexible style modification on generated dances. (2)~We develop a method to represent and learn dance styles in a latent space spanned by a base of prototypes, which are learnt from real dance videos in an end-to-end manner, and in this framework, the style embedding is computed automatically as a linear combination of such prototype vectors with varying weights to indicate different styles. Then, such latent knowledge in the form of style-related weighting of the prototype vectors is transferred to dance generation so as to guarantee style consistence between the real and generated dance videos. (3)~We establish a large music-to-dance dataset that contains choreographies of three dance styles. (4)~The qualitative and quantitative results validate that the proposed framework generates more diverse and realistic dances in regard to input music.


\section{Related Work}

\subsection{Music-to-dance Generation}

The music-to-dance generation models can be roughly divided into two categories: Retrieval-based methods and learning-based methods. Previous works~\cite{shiratori2006dancing,kim2009perceptually,fan2011example,lee2013music} on dance generation carefully design musical feature extractors, and then the dance motion is selected from the database or locally modified by feature matching. These retrieval-based methods suffer from the lack of machine learning to capture the generic correlation between music and dance and cannot generate novel dances unlimitedly.

Recently,~\cite{alemi2017groovenet,tang2018dance,lee2019dancing,ye2020choreonet} attempt to apply deep generative models to generate more creative dances.~\cite{alemi2017groovenet} establishes the connection between music and dance motions through the Factored Conditional Restricted Boltzmann Machines (FCRBM) and Recurrent Neural Networks (RNN) models.~\cite{yalta2019weakly,tang2018dance} design alternative LSTM Auto-Encoder models. The former method concatenates the previous motion with current hidden states to reduce the feedback error. The latter applies the music beat feature as mask to weight the hidden states only to reduce the feature dimensionality as well as the likelihood of overfitting.~\cite{ye2020choreonet,lee2019dancing} decompose the dance into dance units conditioned on the music beat and let the models learn to recompose them according to input music. 

The aforementioned works ignore the knowledge in dance motion referred to as style in this study. Here, we propose a dance generation model with controllable style embedding to apply the latent knowledge learnt from specific-style dances.

\subsection{Realistic Video Generation from Skeletons}

We introduce a rendering model to generate realistic dance videos from kinematic representations of human body. The existing video generation works utilize conditioned generative adversarial networks~(CGAN)~\cite{liu2019liquid,ma2017pose,chan2019everybody,ma2018disentangled,ren2020deep} or Variational Auto-Encoder~\cite{dong2018soft}.~\cite{ma2017pose} proposes a U-Net generator with a coarse-to-fine strategy to generate $256 \times 256$ images.~\cite{liu2019liquid} utilizes Liquid Warping Block (LWB) in body mesh recovery and flow composition, and a GAN module to tackle different tasks.~\cite{chan2019everybody} uses Pix2PixHD~\cite{wang2018pix2pixHD} and special Face GAN to achieve body imitation for generating a more realistic target video. In this work, we simplify~\cite{chan2019everybody} for realistic dance video generation.

\begin{figure*}[t]
\begin{center}
	\includegraphics[width=\linewidth]{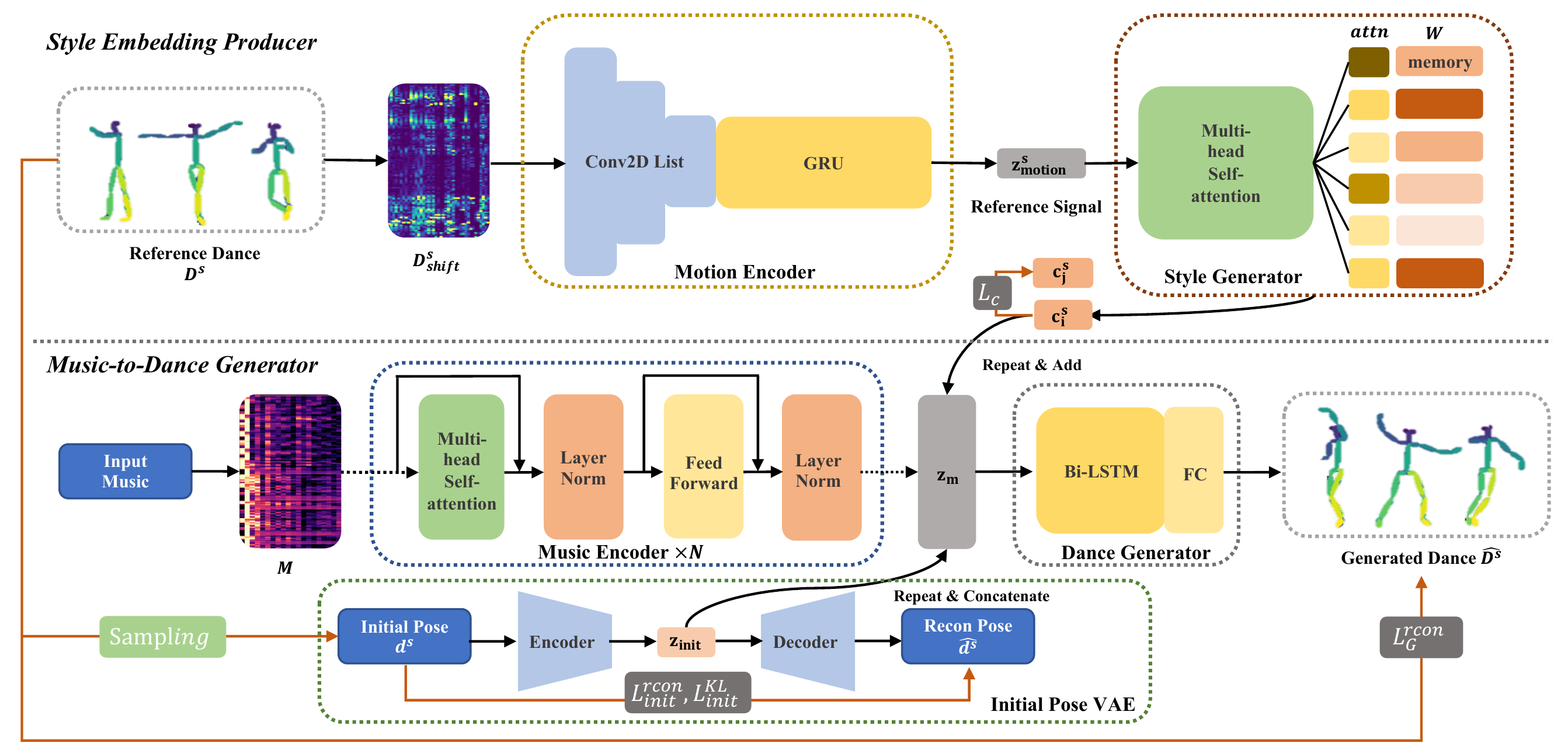}
\end{center}
   \caption{The architecture of controllable style embedding dance generation framework. The framework consists of two parts: A style embedding producer and a music-to-dance generator.}
\label{fig:2}
\end{figure*}

\section{Dataset and Preprocessing}


We build a large music-to-dance dataset with three popular representative categories: Popping dance, anime dance, and locking dance. We do not choose traditional dance styles due to the limited movements. All these videos in the dataset are uploaded onto web by professional dancers, with clear high-quality music, and solo-dance. The Dance videos are resized to $640 \times 480$ to extract the musical features and motion features. Our novel dataset containing 70 popping dance videos, 150 anime dance videos, and 65 locking dance videos, with $1.48M$ carefully selected frames.

{\bf Musical Features Extraction.}~We first normalize the music volume according to root mean square by FFMPEG. Then, the musical features are extracted via an audio analysis library Librosa~\cite{mcfee2015librosa}, including  20-dimensional Mel Frequency Cepstral Coefficients (MFCC), 12-dimensional constant-Q transform chromagram, 1-dimensional pitch, 1-dimensional root-mean-square energy~(RMSE), and 1-dimensional onset strength.

{\bf Motion Features Acquisition.}~All raw videos are fed into Openpose~\cite{cao2017realtime} for 2D keypoints detection. To ensure that the skeleton information in the dance motion sequences is of high quality, we manually filter out the video clips whose skeletons are not detected correctly. We finally choose 21 joints most relevant to kinematic expressiveness to represent the dancing motion, including nose, neck, mid hip, left and right ears, shoulders, elbows, wrists, hips, knees, hand, and ankles.

 \section{Methodologies}

Our goal is to generate multi-style dances conditioned on the input musical features, and initial poses. The style of a sequence of generated dance should be adhered to the style learnt from the reference dance clips. We define the input musical features as $\{M_{t}\in{\mathbb{R}^{L}}|t=1,\dots,T\}$, and $L=36$ is the dimension of the musical features of a single frame. Let $J$ be the number of body joints, each of which is represented by its pixel coordinates $(x, y)$. Hence, a dance motion sequence is formulated as $\{D_t^s \in {\mathbb{R}^{2J}| t=1,\dots,T}\}$ with a style indicator $s\in\{ anime\_dance,  popping\_dance, locking\_dance \}$, where $J=21$ indicates all joint locations.

 As illustrated in Figure~\ref{fig:2}, we propose a novel end-to-end dance generation framework, which consists of a style embedding producer to learn the style information from reference dance clips in a latent manner, and a music-to-dance generation model to map musical features to the dance motions. In the style transfer phase, the style embedding producer acts to summarize the attributes of reference dance motions into a style embedding in the form of a vector indicating the category of dance of interest, which can be regarded as a latent knowledge with varying numerical values to alter style from slightly to remarkably. Then, the music encoder in the music-to-dance generator compresses the musical features into semantic representations. The dance generator final generates the dances conditioned on the music representation, the style embedding, and the initial pose in the forms of a hidden vector resulting from an initial pose VAE model~\cite{lee2019dancing}. In the inference phase, for a specific-style dance generation, we can easily apply a style embedding to the music-to-dance generation model and alter the values of the style embedding to control the style so as to enable diversity of styles. Under the control of style embedding, the music-to-dance generation model can generate dances with any expected style. In the following, we first present the music-to-dance generation model with style embedding. Then, the mechanism of style transfer via style embedding is described. Finally, we demonstrate a new loss that functions to force the learnt representations of style embeddings fall into discriminative clusters of different styles.

\subsection{Music-to-Dance Generation}

Since the relationship between music and dance is frame-aligned, we provide a music-to-dance generator module in an Auto-Encoder architecture to generate dance in a frame-by-frame way. As descripted in Figure~\ref{fig:2}, we apply a set of transformer music encoders with a Bi-LSTM dance generator in music-to-dance generator.

Taking advantage of the special temporal structure composed of fully connected layers, the music encoder can obtain a long-range temporal representation from the input musical features $ M \in {\mathbb{R}^{T \times L}}$. For notational simplicity, we ignore the temporal indicator $t$ in the following. Aside from that, supported by the multi-head attention mechanism, the music encoder can capture different musical semantic information at the distinct heads of attention, which can help dance generator better understand the music. The final output of the encoders is the music representation $z_m \in \mathbb{R}^{T \times d_{model}}$, where $d_{model}$ is the dimension of the output of music encoder.

To facilitate the long-term sequential generation, i.e.,~the last pose of a current dance sequence can be used as the initial pose of the next one, where we embed the dance motions into a latent distribution $Z_{init} \sim \mathcal{N}(0,I)$ via initial pose VAE~(green dashed part in Figure~\ref{fig:2}). We train this model on the dance poses $d^s$ that are sampled from the reference dance $D^s$. We enforce a reconstruction L1 loss $\mathcal{L}_{init}^{rcon}$ and a KL loss $\mathcal{L}_{init}^{KL}$ to enable the reconstruction after encoding and decoding:

\begin{eqnarray}
	\mathcal{L}_{init}^{rcon} = |\hat{d^s} - {d^s}|,  \nonumber \\
	\mathcal{L}_{init}^{KL} = KL(\mathcal{N}(0,I)\|z_{init}),
\end{eqnarray}
where $KL(u\|v)=-\int{u(z)\log\frac{u(z)}{v(z)}}dz$. Utilizing the initial pose VAE, we can randomly sample initial poses and encode the last pose of a current dance clip for the next dance clip generation to enable long-term dance sequential generation at inference stage.

After all, the dance generator receives a music representation $z_m$ from music encoder, a hidden vector $z_{init}$ for an initial pose from initial pose VAE, a style embedding $c^s$ from style generator which will be mentioned in section ~\ref{section:4.2}. We repeat both style embedding $c^s$ and the hidden vector $z_{init}$ of the initial pose at the time dimension to align with the shape of $z_m$. Then we add the expanded style embedding to music representation, and concat with the expanded hidden vector of the initial pose. The composite representation will be passed to the dance generator module for dance generation. Finally, we supervise the generated dance $\hat{D^s}$ by using reconstruction loss:
\begin{equation}
	\mathcal{L}_{G}^{rcon} = |\hat{D^s} - {D^s}|.
\end{equation}

\subsection{Style Embedding Producer}\label{section:4.2}
In fact, human's dance style knowledge is formed in the process of continuous practice and perception of dance motions. Such memory of dance styles as a complex pattern is in general the composition of prototype patterns. Motivated by this, we represent the style embedding as a linear combination of a couple of prototype patterns. The prototype patterns are learnt from the whole corpus of real dance videos and one solution of the weights of such prototype patterns in the linear combination of them indicates a specific dance style for a given video example. Ideally, a remarkable variation of the weights will cause style change while a slight one will not. In the following, we present how to learn the prototype patterns and represent style embedding through the weighted sum of such prototypes in an end-to-end manner. 

As illustrated in Figure~\ref{fig:2}, the style embedding producer first computes the first-order difference between two sequential motion frames as dynamic motion feature: $D_{shift}^s = |D_{t+1}^s - D_t^s|, t = 1,\dots, T$, where $t$ indicates the time step, and $s$ is the label of the corresponding dance style. In the motion encoder, a stack of 2-D convolutional layers and a GRU layer map the dynamic motion feature $D_{shift}^s$ into a fixed-length motion representation vector $z_{motion}^{s} \in \mathbb{R}^{1 \times d_{gru}}$, which is the last state of the GRU layer, where $d_{gru}$ is the hidden size of the GRU layer and the setting of $d_{gru}$ should be equal to that of $d_{model}$ in the music-to-dance generation module. Thus, $z_{motion}^s$ encodes the reference dance motion's spatial and temporal features locally, serving as the reference signal fed to the style generator.

Then, we design a learnable matrix referred to as style memory to learn the prototype patterns from the motion representations globally. As shown in Figure~\ref{fig:2}, we feed the dance clips $D^s$ to the style embedding producer to extract the motion representation $z_{motion}^s$ as the reference signal for guiding the dance style memory $W \in \mathbb{R}^{d_w \times d_{gru}}$ writes in the training phase and reads $W$ in the inference phase, where $d_w$ is a hyper-parameter. Here, each row of $W$ can be regarded as a prototype vector to span a latent space in order to represent any given signal corresponding with dance moves using $d_w$ base vectors, where both the prototype vectors and the weights to form the linear combination of such prototype patterns are learnt in an end-to-end manner. By means of $W$, any given motion signal can be represented as a linear combination of the $d_w$ prototype vectors stored in $W$. In terms of inference, we leverage the multi-head self-attention module to calculate the similarity between the reference signal $z_{motion}^s$ and each prototype row vector in $W$ to represent the given signal in the form of a linear combination of the prototype vectors according to how much it is close to each prototype vector. The attention module outputs $attn \in \mathbb{R}^{1 \times d_w}$ automatically under predefined loss function, which functions to weight the contribution of each prototype style pattern. The weighted sum of such prototype vectors, namely style embedding, is passed to the music-to-dance generator module as conditioning at every time step. So, the style embedding $c^s \in \mathbb{R}^{1 \times d_{gru}}$ is calculated as follow:

\begin{eqnarray}
attn = \frac{Softmax(z_{motion}^sW^T)} {\sqrt{{d_{gru}}}},  \nonumber \\
c^s = Sum(attnW).
\end{eqnarray}
Overall, we jointly train the style embedding producer and music-to-dance generation model by optimizing the following objective:

\begin{equation}
	\mathcal{L} = \mathcal{L}_{G}^{rcon} + \lambda_{init}^{rcon}\mathcal{L}_{init}^{rcon} + \lambda_{init}^{KL}\mathcal{L}_{init}^{KL},
\end{equation}
where $\lambda_{init}^{rcon}$,$\lambda_{init}^{KL}$ are the weights of related loss terms.

\subsection{Learning Representations Converging to Style-related Clusters via a Contractive Loss}

In the learning of the prototype patterns and the weight sum of them to form the style embedding, the loss function plays an important role. A proper loss function makes the leant representations of dance styles fall into compact and separable clusters while a bad one will cause overlap of the representations over different style categories. In the following, we first visualize how a bad one works, and then, how a proper one works.

\begin{figure}[t]
\begin{center}
	\includegraphics[width=\linewidth]{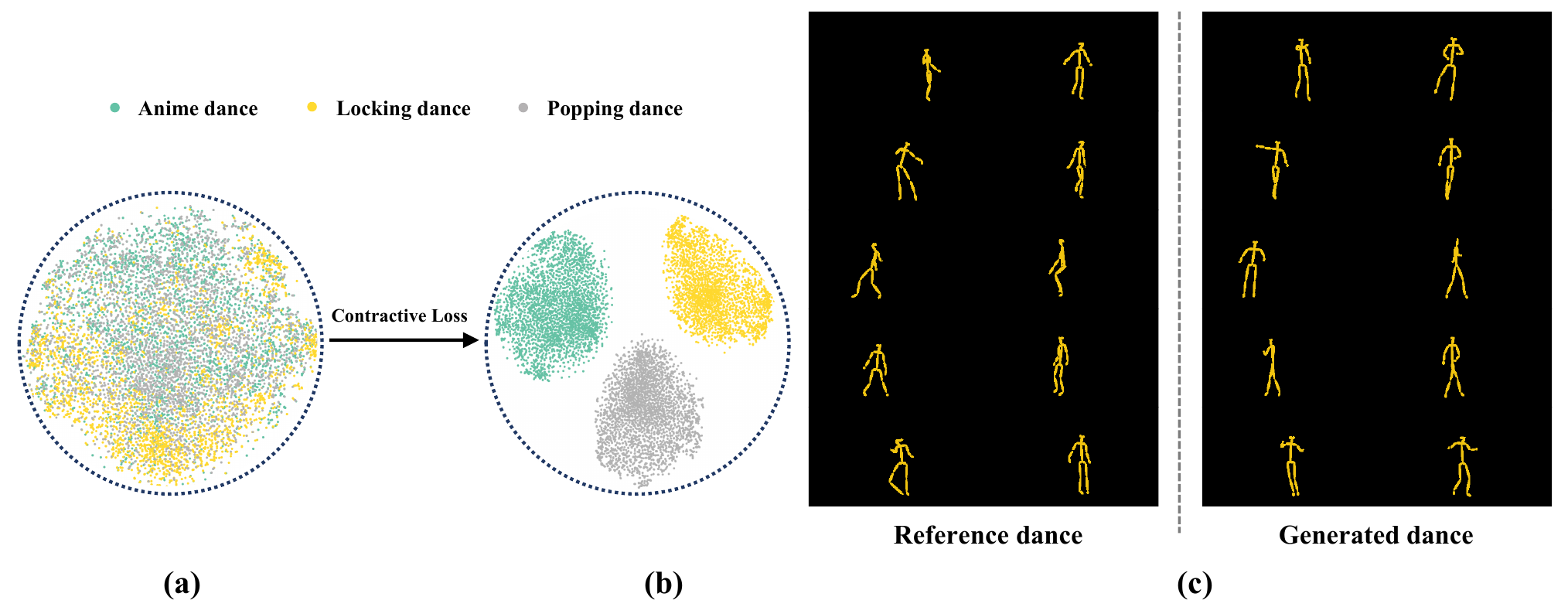}
\end{center}
   \caption{The visualization of the PCA of style embeddings, an example of reference dance clip, and the generated dance clip: (a)~The PCA of style embeddings without style label supervision. (b)~The PCA of style embed-dings with style label supervision. (c)~At left is the reference dance clip used to produce style embedding for dance generation in the framework with supervision of style labels. At right is the dance clip generated with the style embedding produced by the reference dance clip at left.}
\label{fig:3}
\end{figure}

To enable visualization of the high-dimensional style embedding $\{C^s\}$ of all the examples, we reduce the style embeddings' dimension with principal components analysis~(PCA) to observe the relationship between the style embeddings resulting from inappropriate loss function and their labels of style categories. As shown in Figure~\ref{fig:3}~(a), confusion among different style categories is obvious. To alleviate that, we develop a contrastive loss $\mathcal{L}_c$ to introduce weak label information to the style embedding producer to force the learnt representations of styles $\{C^s\}$ fall into separable clusters corresponding with different style categories. The contractive loss $\mathcal{L}_{c}$ is defined as follows:

\begin{eqnarray}
\mathcal{L}_{c}(c_i^s, c_j^s) = \frac{1}{2}(1-Y)(Dis)^2 + \nonumber \\
 \frac{1}{2}(Y)\{max(0, m-Dis)\}^2,
\end{eqnarray}
where the $(c_i^s, c_j^s)$ is a pair of style embeddings, and the $i, j$ are just used to identify the different style embeddings. If the pair of style embeddings $(c_i^s, c_j^s)$ are generated from the reference dances of an identical style, $Y=0$. Otherwise $Y=1$. $Dis$ is the Euclidean distance between $c_i^s$ and $c_j^s$. $m$ is margin. We set it to $1$ to separate different style samples. Thus, the overall objective $\mathcal{L}$ changes to:
\begin{equation}
	\mathcal{L} = \mathcal{L}_{G}^{rcon} + \lambda_{init}^{rcon}\mathcal{L}_{init}^{rcon} + \lambda_{init}^{KL}\mathcal{L}_{init}^{KL} + \lambda_{style}\mathcal{L}_{c},
\end{equation}
where $\lambda_{style}$ is the weight of related loss terms.

As a result, Figure~\ref{fig:3}~(b) illustrates that the style embeddings obtained under such loss function fall into different clusters with a strong correlation to the style labels, which enable us to perform style control through style embedding. Finally, we visualize an example of the generated dance clip corresponding to the style embedding calculated from the example reference dance in Figure~\ref{fig:3}~(c). As indicated in Figure~\ref{fig:3}~(c), though the generated dance and reference dance are totally different choreographies, the moving direction~(move left then right) of the generated dance is same as that of the reference dance, and both the generated and the reference dance have the same rotation behaviors.~(More cases displayed in {\bfseries supplementary materials} show that the style embeddings can capture distinct attributes of dance motions, such as moving speed, moving direction, and appearance, which are commonly used to characterize the style of dance. Based on that, we can successfully control the styles of the generated dances by the style embeddings lerant from the input reference dance clips.)

 \section{Experiments}
 To evaluate our method, we conduct extensive experiments on our music-to-dance dataset. Our experiments include both quantitative and qualitative evaluations. We compared our method with the LSTM method proposed in~\cite{shlizerman2018audio}, MelodyNet~\cite{tang2018dance}, and the state-of-art method Dancing2Music~\cite{lee2019dancing}. More qualitative evaluation results and the implementation are detailed in {\bfseries supplementary materials}.


\begin{table}[b]
\centering
\scalebox{0.9}{
\begin{tabular}{lccccc}
\toprule
Method   & Beat Hit  & Style Consistency &  Diversity \\
\midrule
Real Dance       & 63.6\%         & 2.5        & -			     \\
\midrule
   LSTM              & 21\%           & 5.3        & -               \\
   MelodyNet         & 49\%           & 3.6        & -               \\
   Dancing2Music     & 68.5\%         & 2.9        & 2.5              \\
   Ours              & {\bf69.8\%}    & {\bf2.65}  & {\bf2.9}         \\
\bottomrule
\end{tabular}}
\caption{The quantitative evaluation results.}
\label{tab:2}
\end{table}

\subsection{Quantitative Evaluation}

{\bf Beat and intensity.}~Whether the beat of dance matches music beat greatly affects the quality of the generated dance. Given all the music and generated dances, we gather the music beats $B_m$, the number of total dance beats $B_d$, the number of dance beats that are aligned with music beats noted as $B_a$. We use $B_a / B_m$ as the beat hit rate to evaluate how generated dance match the rhythm of music. The music beat information $B_m$ can be easily extracted from music onset features by Librosa. For dance, we regard the frames that the movement drastically slows down as a dance beat event. In practice, we compute the motion magnitude between neighboring dance motions, and track the magnitude trajectories to locate when a dramatic decrease in the motion magnitude appears. As shown in Table~\ref{tab:2}, our method outperforms Dancing2Music by 1.3\% in terms of beat hit metric, due to the frame-by-frame transformer module to align the dance motions with the musical features.

\begin{figure}[t]
\begin{center}
	\includegraphics[width=\linewidth]{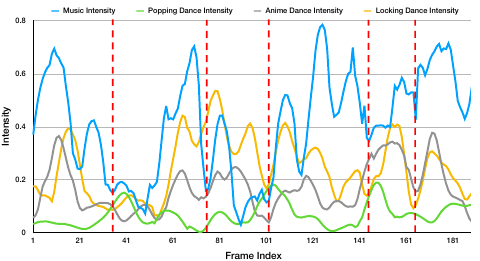}
\end{center}
   \caption{Intensity of the music and generated dance.}
\label{fig:4}
\end{figure}

 We also introduce the intensity metric to evaluate the generated dances in terms of how much the emotion of music has been exhibited by dance. For music, the RMSE feature can represent the intensity of the music. For dance, we apply the sliding window to average over the magnitudes of dance motions as the intensity of the dance. As shown in Figure~\ref{fig:4}, we draw the intensity curves to describe the emotion of the music and the generated dance. We find that the generated dance responses timely to the input music in terms of emotion, as the dance moves will be more energetic when the music appears to be more intense, and on the contrary, the moves will be more slow and deliberate.

\begin{figure}[b]
\begin{center}
	\includegraphics[width=\linewidth]{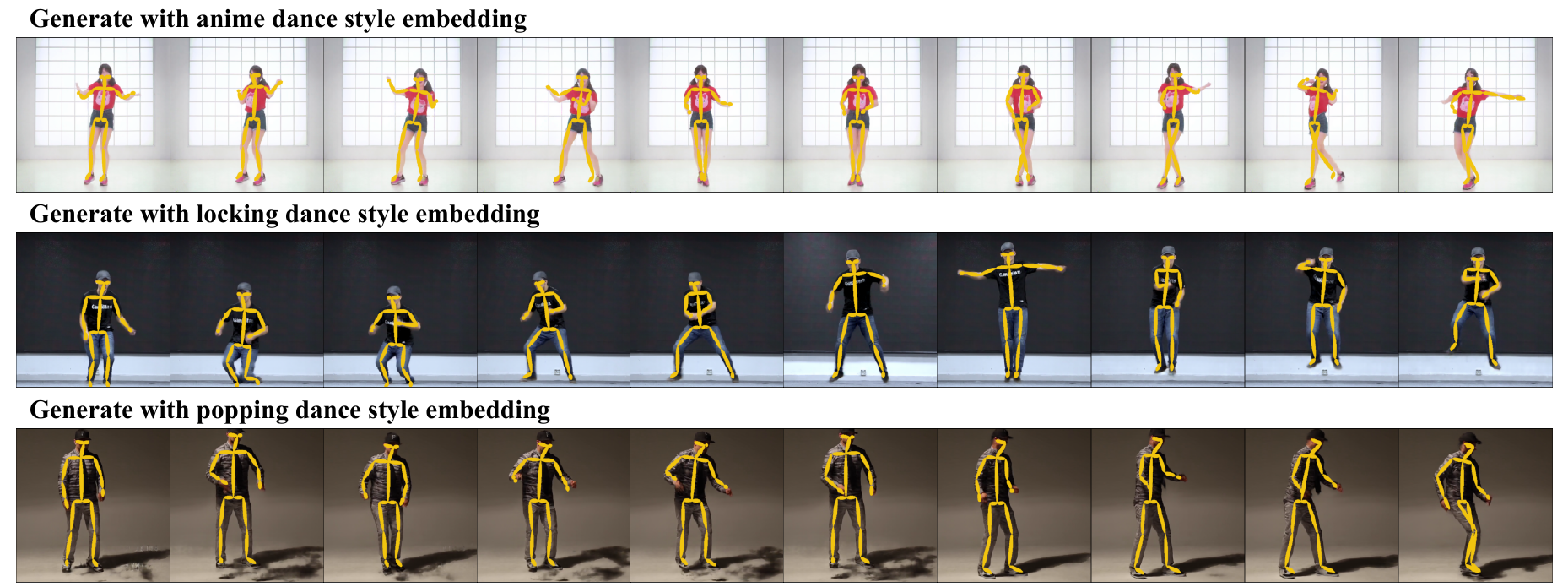}
\end{center}
   \caption{The visualization of the generated dances with our render models.}
\label{fig:5}
\end{figure}

\begin{figure}[t]
\begin{center}
	\includegraphics[width=\linewidth]{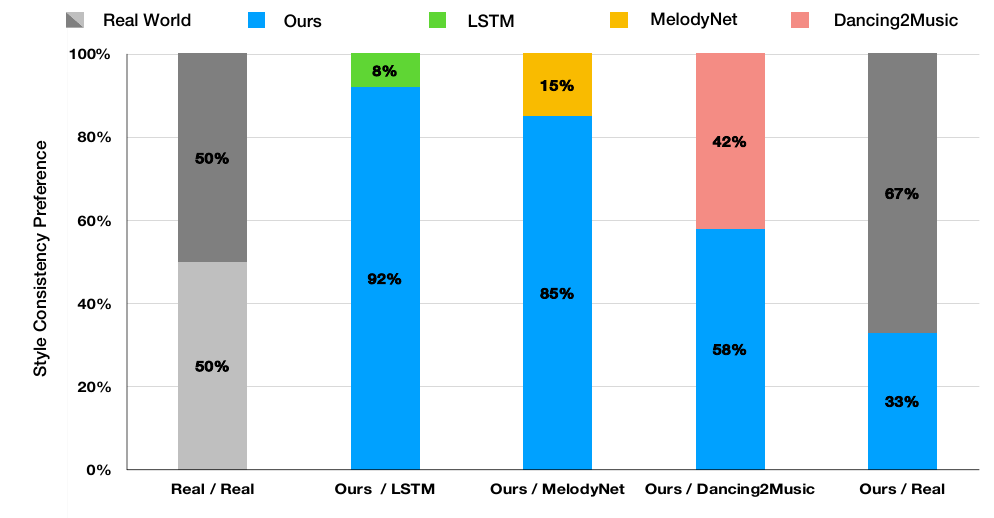}
\end{center}
   \caption{The preference results on style consistency.}
\label{fig:6}
\end{figure}

{\bf Style Consistency and Dance Diversity.} We use the Fréchet Inception Distance~(FID) to estimate the style consistency for the generated dances with the same styles. The FID score is also used to measure the diversity of the generated dances conditional on the same music but with different initial poses. To evaluate the style consistency, we extract the style embeddings of the synthesized dances and those of the training dances. As depicted in Table~\ref{tab:2}, our style consistency score is lower than those of baselines which means our model can generate the precise style of dances. We also attempt to generate dances on the 20 test music clips with 50 randomly-picked initial poses and calculate the FID score between the style embeddings of generated dances as diversity score. The results in Table~\ref{tab:2} show that our model is significantly better than the Dancing2Music method in terms of diversity score, which means more generative. The reason is that the generated dances from Dancing2Music are based on a global music style representations from the music classifier while the same style music have the almost same style features. In contrast, our model considers the frame-level musical features that are strictly aligned with the dance motion frames according to different weights from the transformer encoder.

\subsection{Qualitative Evaluation}

{\bf Style Transfer with Controllable Style Embeddings.} To show that our model can generate accurate styles of dances corresponding to the style embeddings. we generate three style embeddings with an anime dance clip, a locking dance clip, and a popping dance clip. Then, we put these embeddings to the dance generation model and generate dances with the same music and initial pose. The output dances are converted into real videos via the rendering models. The rendered results are shown in Figure~\ref{fig:5}, in which we can find that the model can produce a corresponding style of dance following style embedding. More results are described in the {\bfseries supplementary material}.

 {\bf User Study on Style Consistency.} We conduct a user study using a pairwise comparison scheme to verify the performance of our music-to-dance framework in style consistency. Given a pair of dances generated from the different models which are required to generate specific-style dances, the users are asked to select which one is closer to the requested dance style. Figure~\ref{fig:6} shows the user study results, where our approach outperforms the baselines in style consistency. However, compared with the real dance, there is still a gap. The distortions of limbs in some special cases make the generated dances unnatural.

\section{Conclusion}
In this work, we propose a novel style controllable dance generation framework that can learn different style patterns from the reference dances and transfer them to the generated dances in end-to-end manner. We also introduce a new large-scale music-to-dance dataset. Combing the dance generation framework with a rendering model, we create a successful cross-modal generation framework that can directly generate dance videos from music. It is the first work to adapt the style of real dance videos to generated videos. In future works, instead of joint coordinates, we will attempt to design a dance generation network directly based on the dance video frames and expand the framework to generate multi-people dances.

\clearpage

\bibliographystyle{named}
\bibliography{ijcai21}

\end{document}